\documentclass{epl}
\usepackage{color}

\begin{document}

\title{Orbital ordering in the two-dimensional ferromagnetic 
       semiconductor $ {\bf Rb_2CrCl_4} $}

\shorttitle{Orbital ordering in $ {\rm Rb_2CrCl_4} $}

\author{U.\ Schwingenschl\"ogl
        \and V.\ Eyert
        \thanks{E-mail:\ \email{Volker.Eyert@physik.uni-augsburg.de}}}

\institute{Institut f\"ur Physik, Universit\"at Augsburg, 
           86135 Augsburg, Germany}

\pacs{71.20.-b}{Electron density of states and band structure of 
                crystalline solids}
\pacs{71.20.Be}{Transition metals and alloys}

\maketitle

\begin{abstract}
We present the results of electronic structure calculations for the 
two-dimensional ferromagnet $ {\rm Rb_2CrCl_4} $. They are obtained 
by the augmented spherical wave method as based on density functional 
theory and the local density approximation. In agreement with
experimental data $ {\rm Rb_2CrCl_4} $ is found to be semiconducting 
and displays long-range ferromagnetic order of the localized Cr $ 3d $ 
moments. The magnetic properties are almost independent of the 
structural modifications arising from the Jahn-Teller instability, 
which leads from the parent body-centered tetragonal $ {\rm K_2NiF_4} $ 
structure to a side-centered orthorhombic lattice. In contrast, our 
calculations give evidence for a strong response of the optical band 
gap to the corresponding structural changes. 
\end{abstract}

\section{Introduction} 

Since the discovery of $ {\rm CrBr_3} $ as a ferromagnetic semiconductor  
by Tsubokawa \cite{tsubokawa60} this narrow class of materials has 
attracted increasing interest. To some part this is for fundamental 
reasons, since these compounds serve as ideal candidates for the 
Heisenberg model. However, the prospect of using ferromagnetic 
semiconductors in spin electronics has motivated an even greater amount 
of research activities in the last years. Only recently, 
$ {\rm La_2MnNiO_6} $ has been shown to exhibit long-range 
ferromagnetic order below $ {\rm T_C} \approx 280 $\,K with a 
magnetization of 5\,$ \mu_B $ per formula unit \cite{rogado05} 
and an optical band gap of about 0.6\,eV as determined by electronic 
structure calculations \cite{matar05}. 

Rubidium tetrachlorochromide, $ {\rm Rb_2CrCl_4} $, also belongs to 
this class of non-metallic ferromagnets albeit with a much lower 
ordering temperature of $ {\rm T_C} = 52.4 $\,K 
\cite{gregson75,hutchings81}. According to neutron scattering 
measurements of the spin-wave dispersion this compound behaves as 
a two-dimensional, $ S = 2 $, easy-plane ferromagnet with 
an in-plane exchange constant of $ J = 7.7 $\,K \cite{hutchings76}. 
Since the interlayer coupling was estimated to be about four 
orders of magnitude smaller than the intralayer interaction, 
three-dimensional long-range order was also attributed to the 
weak uniaxial anisotropy. The latter, in addition, leads to a 
small energy gap of about 0.1\,meV in the spin wave dispersion 
at the center of the Brillouin zone, which precludes true XY-type 
behaviour \cite{hutchings81,janke83}. Furthermore, the anisotropy 
causes a canting of the ordered spins of $ 5.2^{\circ} $ away from the 
$ \langle110\rangle $ direction \cite{janke83}. From susceptibility 
measurements and neutron scattering investigations of the static critical 
properties a crossover from an ideal two-dimensional XY-like behaviour 
to a narrow three-dimensional region on approaching $ {\rm T_C} $ from 
either side of the transition was inferred 
\cite{kleemann86,alsnielsen93,bramwell95}.

The crystal structure of $ {\rm Rb_2CrCl_4} $ is closely related 
to the body-centered tetragonal structure of $ {\rm K_2NiF_4} $ 
with space group $ D_{4h}^{17} $ (I4/mmm). In this structure the 
magnetic ions are arranged on a square planar lattice and coupled 
via the anions located midway between them. However, the Cr $ d^4 $ 
electronic configuration in $ {\rm Rb_2CrCl_4} $ with triply occupied 
$ t_{2g} $ states and an orbital degeneracy in the singly occupied 
$ e_g $ states is susceptible to a Jahn-Teller instability. Indeed, 
the situation is not unlike that found for  $ {\rm K_2CuF_4} $, which 
has a Cu $ d^9 $ configuration. While for the copper compound early 
experiments were interpreted in favour of the $ {\rm K_2NiF_4} $ 
structure with octahedra compressed parallel to the $ c $-axis, 
Khomskii and Kugel argued on the basis of theoretical considerations that 
this structure was incompatible with the observed ferromagnetic order 
\cite{khomskii73,kugel82}. Subsequently performed neutron and x-ray 
diffraction studies confirmed these ideas and gave evidence for a 
side-centered orthorhombic lattice with space group $ D_{2h}^{18} $ 
(Cmca) and two formula units per cell \cite{ito76,hidaka79}. In 
this structure, the octahedra are elongated with the long axis 
lying in the plane and pointing alternatively parallel to the 
$ a $- and $ b $-axis of the parent $ {\rm K_2NiF_4} $ unit cell. 
For $ {\rm Rb_2CrCl_4} $, first evidence for the orthorhombic 
structure came from NMR experiments, which revealed that the 
local environments of the chromium sites are elongated chlorine 
octahedra having almost tetragonal symmetry \cite{ledang77}. Lateron, 
the space group $ D_{2h}^{18} $ (Cmca) was also confirmed for 
$ {\rm Rb_2CrCl_4} $ \cite{day79,muenninghoff80,janke83}. 

In this space group the Rb atoms occupy the Wyckoff positions (8d), 
$ (0,0,\pm z_{\rm Rb}) $, with $ z_{\rm Rb}=0.3568 $ and the Cr atoms 
are located at the (4a) sites at $ (0,0,0) $. The apical 
Cl atoms are found at the Wyckoff positions (8d) with 
$ z_{\rm Cl}=0.1508 $, whereas the equatorial Cl atoms occupy the 
(8f) positions at $ (1/4+\delta,1/4+\delta,0) $. Here, the distortion 
parameter $ \delta $ characterizes the deviation from the ideal 
$ {\rm K_2NiF_4} $-structure; the experimental value of 
$ \delta=0.0154 $ corresponds to a $ 0.16 $\,\AA\ shift of the Cl 
ions off the central positions between two neighbouring Cr sites 
\cite{janke83}. As a consequence, the in-plane Cr-Cl bond distances 
amount to 2.70 and 2.39\,\AA, whereas the apical distance is 2.37\,\AA. 

Despite their analogous crystal structures there is one important 
difference distinguishing $ {\rm Rb_2CrCl_4} $ from 
$ {\rm K_2CuF_4} $. In the latter compound, the single unpaired 
electron in the $ d^9 $ configuration of the Cu$^{2+}$ ion occupies 
the $ d_{z^2-x^2} $ or $ d_{z^2-y^2} $ orbital, depending on the 
orientation of the local octahedron. Due to the peculiarities of 
the crystal structure the local axes of neighbouring Cu sites are 
orthogonal and so are the orbitals. Hence, according to the 
Goodenough-Kanamori-Anderson (GKA) rules ferromagnetic order results. 
In contrast, the $ d^4 $ configuration of the Cr$^{2+}$ ions is 
characterized by a single $ e_g $ electron in a $ d_{3y^2-r^2} $ 
or $ d_{3x^2-r^2} $ orbital. Yet, despite their $ 90^{\circ} $ 
rotation, orbitals of this kind placed at neighbouring sites are 
not orthogonal and the GKA rules would be in favour of an 
antiferromagnetic coupling. 

In order to resolve this issue in a more systematic manner, 
Feldkemper and Weber started from a multiband Hubbard model, 
which included both the metal $ d $ and the ligand $ p $ 
orbitals \cite{feldkemper98}. Using a Rayleigh-Schr\"odinger 
perturbation expansion to map this model onto an effective 
Heisenberg Hamiltonian they were able to describe the 
superexchange coupling between magnetic sites. Due to the 
inclusion of various interaction paths their calculations 
extended the basic GKA rules and resulted in a net 
ferromagnetic interplanar coupling in agreement with the 
experimental findings. 

In this Letter, we take account of the obvious lack of first 
principles studies for $ {\rm Rb_2CrCl_4} $ and complement the 
model approaches by electronic structure calculations, thereby 
continuing previous work on ferromagnetic semiconductors 
\cite{kuebler88,eyert93,eyert95,eyert97}. In doing so, we aim 
especially at extending the current understanding of the 
interplay between the Jahn-Teller distortion and the optical 
band gap as well as the magnetic ordering. 
Previous electronic structure calculations for $ {\rm K_2CuF_4} $, 
which confirmed the non-metallic ferromagnetic ground state, 
demonstrated, that neither the lattice distortion due to the 
Jahn-Teller instability nor the magnetic order alone are sufficient 
to produce the optical band gap \cite{eyert93}. Since ferromagnetic 
ordering was also found for the ideal $ {\rm K_2NiF_4} $ structure 
($ \delta=0 $), it is the semiconducting ground state, which 
requires the Jahn-Teller effect, and not the magnetic order. Our 
results for $ {\rm Rb_2CrCl_4} $ are in line with this interpretation.

\section{Methodology} 

The calculations are based on density functional theory and the local 
density approximation as implemented in the scalar-relativistic 
augmented spherical wave method \cite{williams79,eyert00b}. In order 
to represent the correct shape of the crystal potential in the large 
voids of the open crystal structure, additional augmentation spheres 
were inserted. Optimal augmentation sphere positions as well as radii 
of all spheres were automatically generated by the sphere geometry 
optimization (SGO) algorithm \cite{eyert98}. Self-consistency was 
achieved by an efficient algorithm for convergence acceleration 
\cite{mixpap}. Brillouin zone sampling was done using an increasing 
number of $ {\bf k} $-points ranging up to 576 points within the
irreducible wedge.

\section{Results and Discussion}

\begin{figure}[htb]
\twoimages[width=65mm]{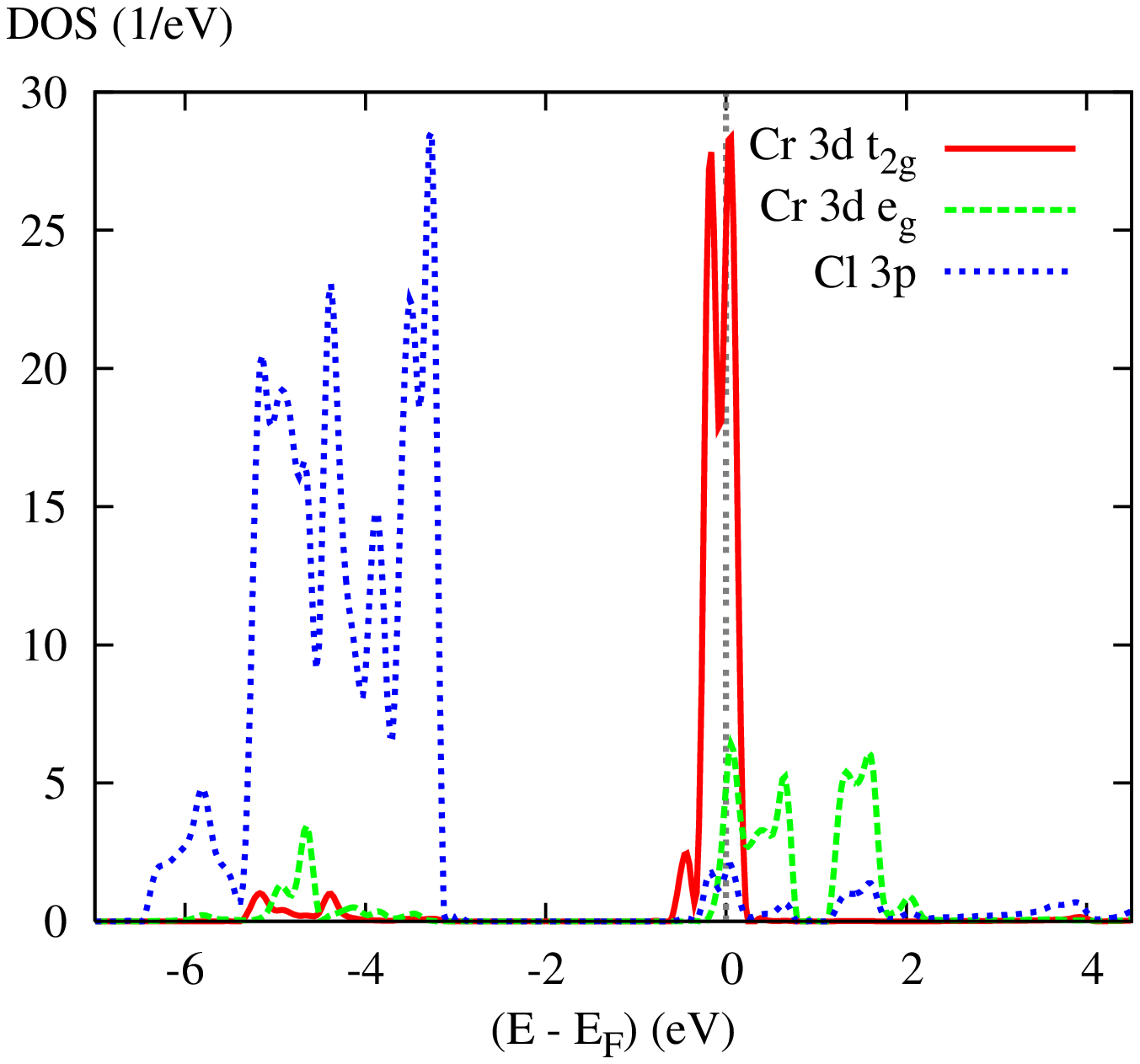}{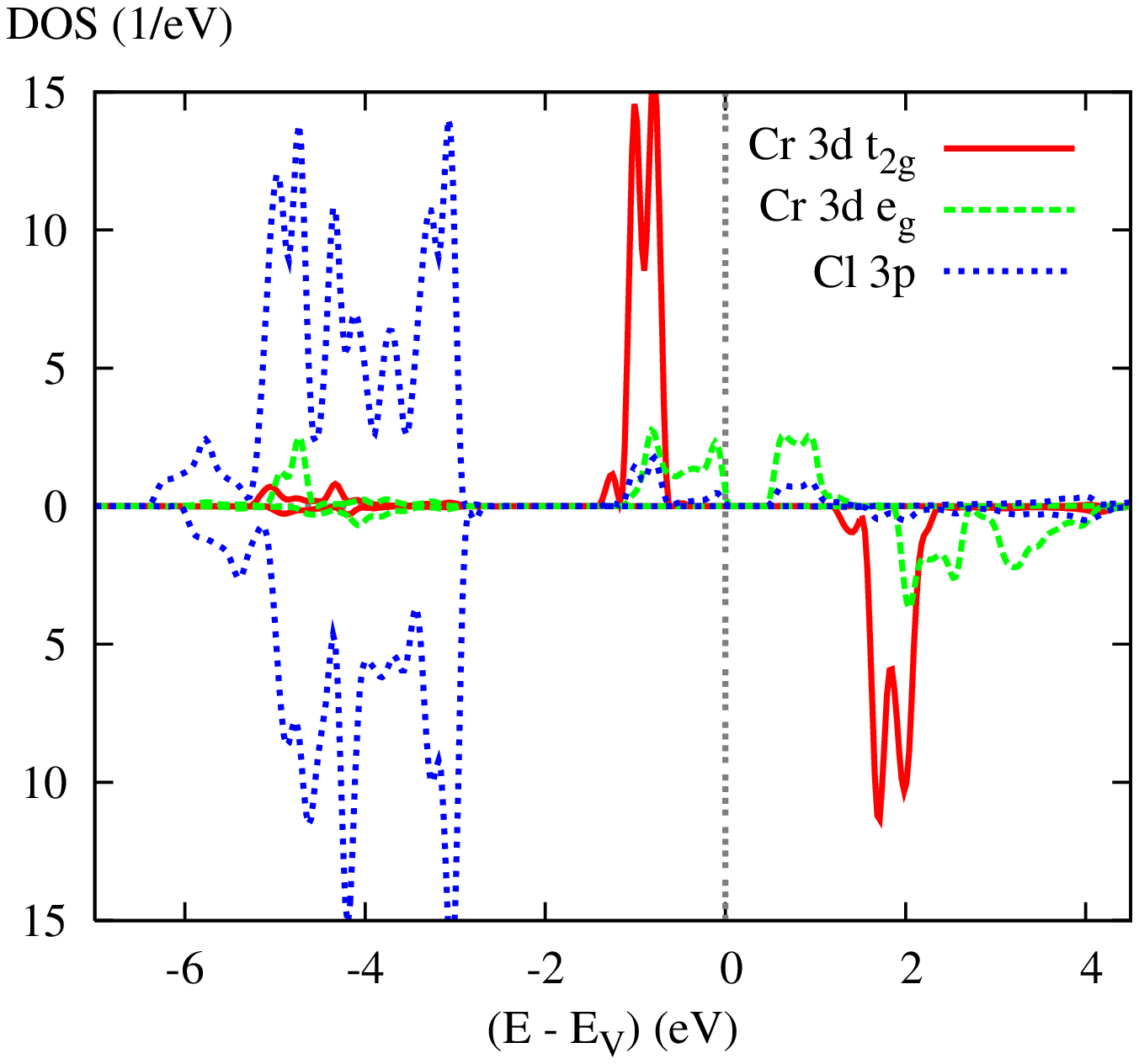}
\caption{Partial Cr $ 3d $ and Cl $ 2p $ DOS for the experimental 
         crystal structure ($ \delta=0.0154 $). Here and in the 
         following figures results of 
         spin-degenerate and spin-polarized calculations are 
         given on the left and right, respectively.}
\label{fig1}
\end{figure}

Partial densities of states as grown out of both spin-degenerate 
and spin-polarized calculations are displayed in Fig.\ \ref{fig1}.  
Of the three groups of bands the first one extending from about 
$ -6.5 $\,eV to $ -3 $\,eV originates predominantly from the 
Cl $ 3p $ states. In contrast, the second and third group as 
found at and above the Fermi level trace back mainly to the 
Cr $ 3d $ $ t_{2g} $ and $ e_g $ states. $ d $-$ p $ hybridization, 
leading to admixtures of these states in the energy region, where 
the respective other partner dominates, is much reduced. Finally, 
taking into account spin-polarization by allowing for long-range 
ferromagnetic order, we observe an almost rigid exchange splitting 
of the $ d $ states of $ \approx 2.8 $\,eV going along with a 
total energy lowering of about 0.3\,Ryd per unit cell. The 
calculated magnetic moment of $ 3.718 \mu_B $ per chromium site 
agrees perfectly well with the value of $ 3.7 \pm 0.4 \mu_B $ 
as deduced from neutron diffraction data \cite{fair77}. 

\begin{figure}[htb]
\twoimages[width=65mm]{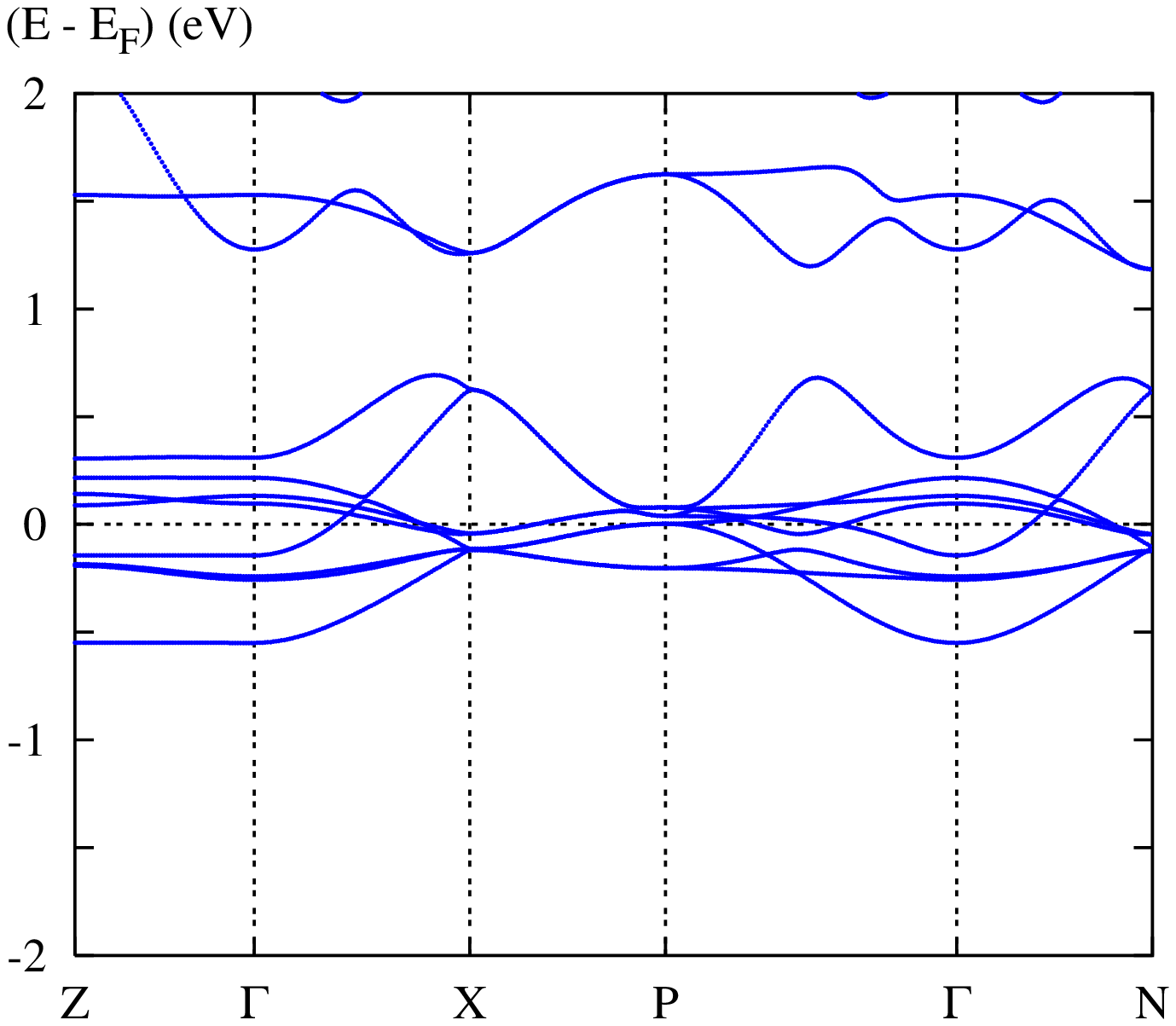}{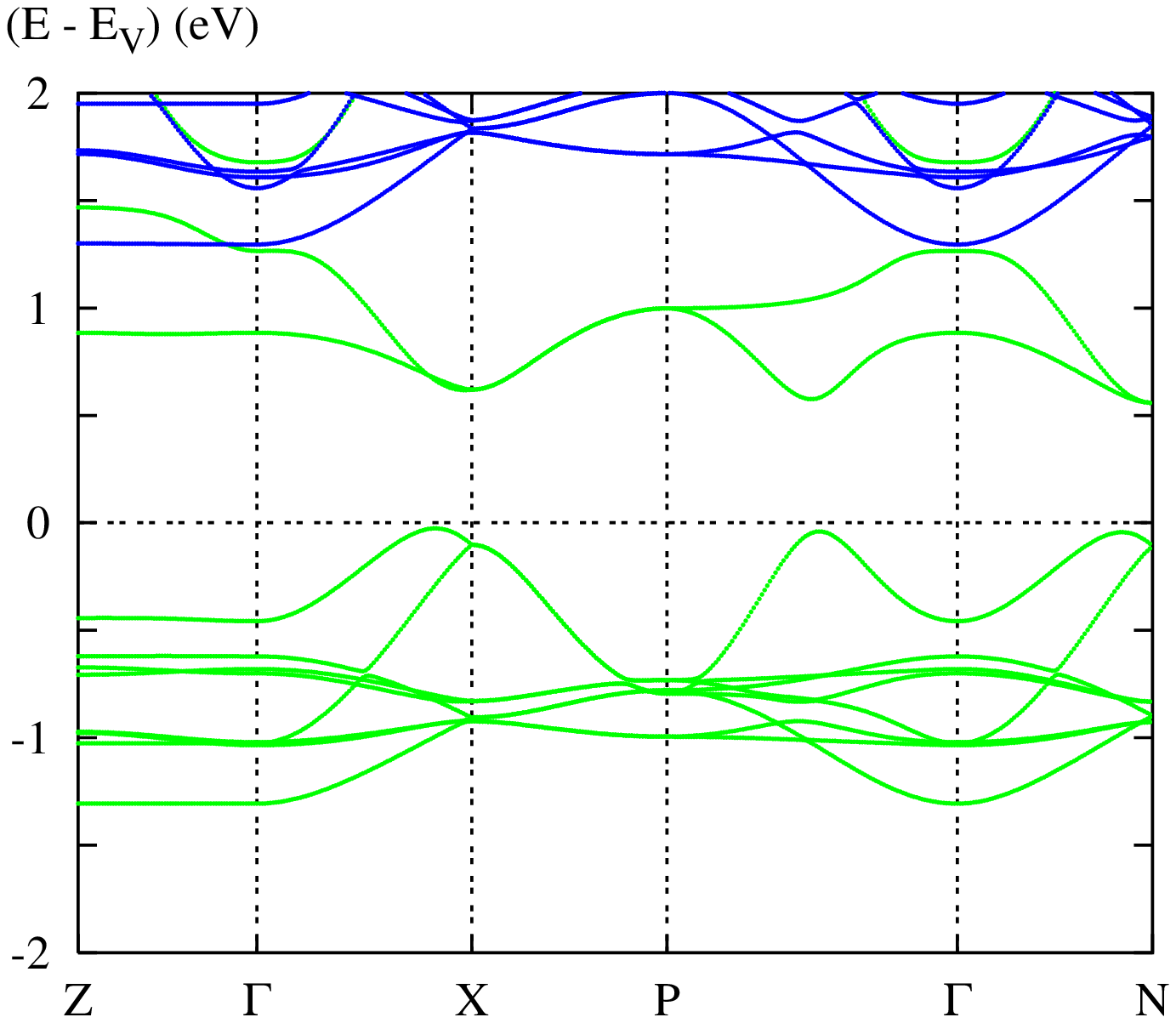}
\caption{Electronic structure for the experimental crystal 
         structure ($ \delta=0.0154 $). In the spin-polarized 
         calculation spin-majority and spin-minority bands 
         are given in green and blue, respectively.} 
\label{fig2}
\end{figure}

\begin{figure}[htb]
\twoimages[width=65mm]{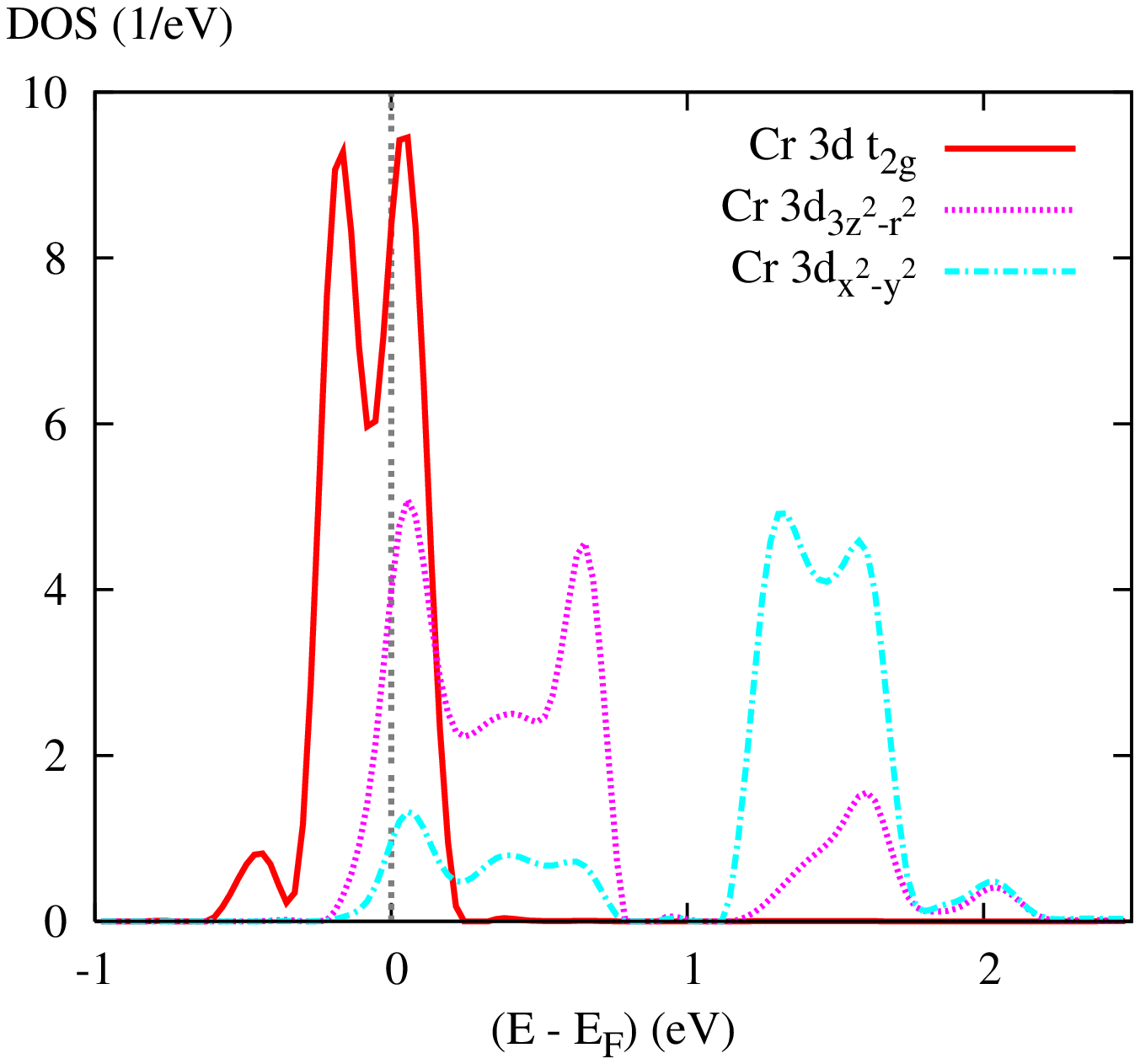}{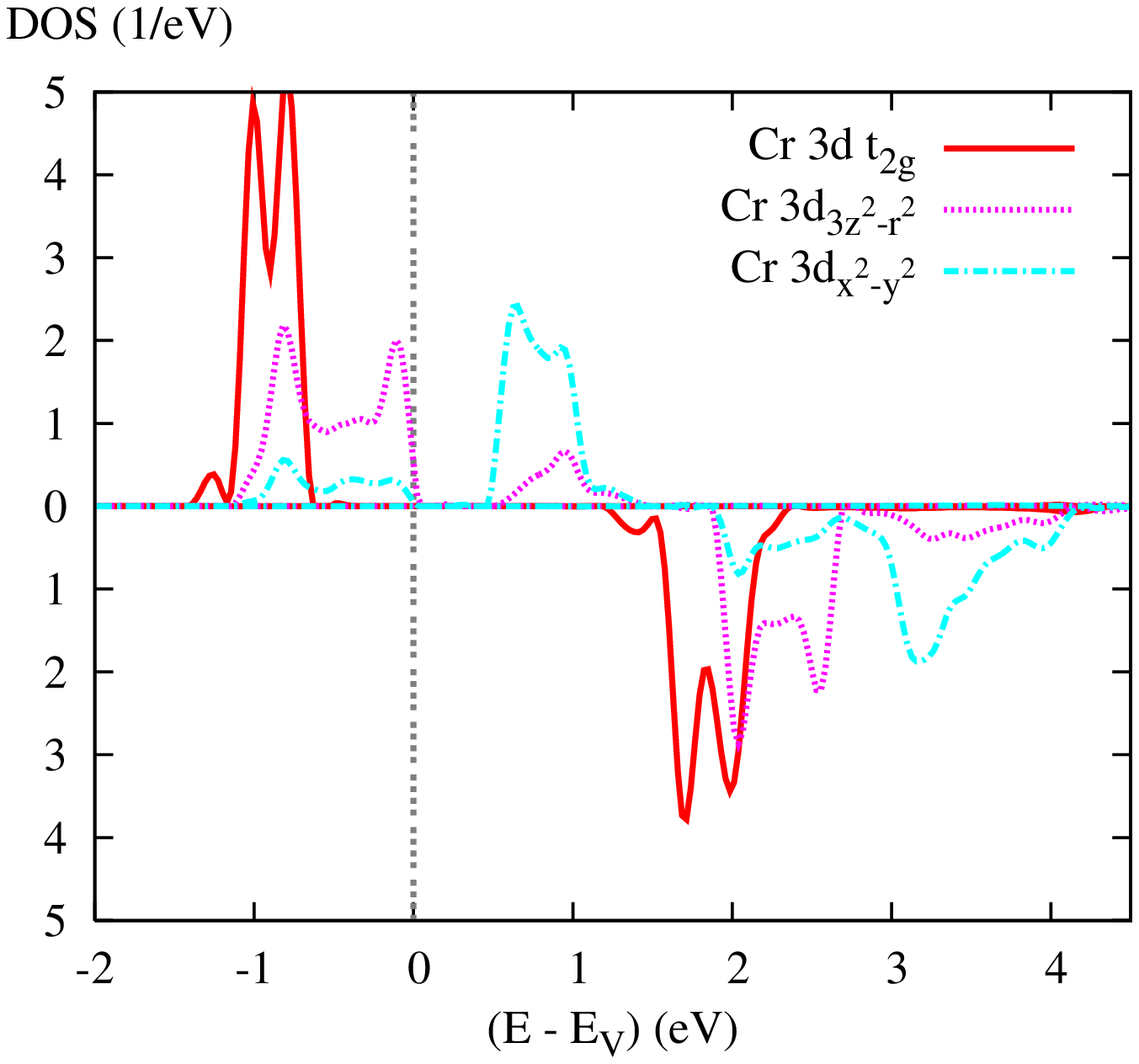}
\caption{Partial Cr $ 3d $ DOS for the experimental crystal 
         structure ($ \delta=0.0154 $). The $ t_{2g} $ 
         contribution is scaled to one orbital.}
\label{fig3}
\end{figure}

Concentrating on the small energy region of the Cr $ 3d $ states 
we display in Figs.\ \ref{fig2} 
and \ref{fig3} 
the electronic band structure and the partial Cr $ 3d $ densities 
of states. For the former we use the Brillouin zone of the 
side-centered orthorhombic lattice with the high-symmetry points 
$ {\rm Z} = (1/2,0,1/2) $, $ {\rm X} = (1/2,0,0) $, 
$ {\rm P} = (1/2,1/2,0) $, and $ {\rm N} = (0,1/2,0) $ (see Figs.\ 
4 and 5 of Ref.\ \cite{eyert93} for more details on the relation 
between the Brillouin zones of $ {\rm K_2NiF_4} $ and 
$ {\rm K_2CuF_4} $/$ {\rm Rb_2CrCl_4} $). 
The almost vanishing dispersion along the line $ \Gamma $-$ Z $ 
reflects the two-dimensional nature of the crystal structure. 
In the representation of the partial densities of states we 
use the local coordinate system with the $ z $ axis pointing 
parallel to the elongated distance of the local $ {\rm CrCl_6} $ 
octahedron. We distinguish the narrow Cr $ 3d $ $ t_{2g} $ states 
from the $ d_{3z^2-r^2} $ and $ d_{x^2-y^2} $ contributions. Due 
to the elongation of the octahedra the $ d_{3z^2-r^2} $ states 
are shifted to lower energies as compared to the $ d_{x^2-y^2} $ 
states. As a consequence, we observe energetical overlap of the 
$ d_{3z^2-r^2} $ states with the $ t_{2g} $ bands and a gap of 
about $ 0.5 $\,eV between the $ e_g $ states. 

While the spin-degenerate calculations lead to metallic behaviour, 
the exchange splitting places the Fermi energy between the 
spin-majority $ d_{3z^2-r^2} $ and $ d_{x^2-y^2} $ states, leaving 
an optical band gap of $ 0.56 $\,eV. Our results thus clearly 
confirm the orbital ordering as deduced from the polarized neutron 
diffraction data by M\"unninghoff {\em et al.}\ with all five 
$ d $ orbitals singly occupied except for the $ d_{x^2-y^2} $ 
orbital, which is empty \cite{muenninghoff82}. 

In addition to the above calculations we have considered an 
antiferromagnetic order with antiparallel spin alignment of 
the two Cr sites inside the side-centered orthorhombic unit cell. 
Again, the bands experience an almost rigid spin splitting, which 
gives room for an optical gap of $ 0.43 $\,eV. The magnetic moments 
per chromium site amount to $ \pm 3.627 \mu_B $. Yet, the energy 
of the antiferromagnetic solution is by $ 27 $\,meV higher than that 
found for the ferromagnetic situation. 

All calculations presented so far were based on the experimental 
crystal structure, i.e.\ for $ \delta=0.0154 $. In order to study 
the effect of the distortion on the electronic and magnetic 
properties we performed additional calculations for several values 
of $ \delta $ varying from zero to 1.5 times the experimental value. 
The results are summarized in Table \ref{tab1},  
\begin{table}[htb]
\begin{tabular}{ccccccc}
Distortion $\delta$ & 
$ {\rm E_{fe}-E_{nm}} $ & $ {\rm \mu_{Cr,fe}} $ & $ {\rm \Delta_{fe}} $ & 
$ {\rm E_{af}-E_{nm}} $ & $ {\rm \mu_{Cr,af}} $ & $ {\rm \Delta_{af}} $ \\
\hline
\hline
 0.0000           & -0.301 & 3.736 &   -   & -0.263 & 3.602 & 0.015  \\
 0.0038           & -0.304 & 3.717 & 0.202 & -0.269 & 3.604 & 0.127  \\
 0.0077           & -0.304 & 3.718 & 0.246 & -0.269 & 3.609 & 0.150  \\
 0.0115           & -0.305 & 3.717 & 0.280 & -0.271 & 3.610 & 0.179  \\
 0.0154           & -0.310 & 3.718 & 0.558 & -0.283 & 3.627 & 0.427  \\
 0.0192           & -0.308 & 3.722 & 0.363 & -0.278 & 3.621 & 0.265  \\
 0.0231           & -0.308 & 3.723 & 0.374 & -0.214 & 3.625 & 0.273  \\
\hline
\end{tabular}
\caption{\rm Total energies (in Ryd per 2 Cr), magnetic moments (in $ \mu_B $),
         and optical band gap (in eV) for different distortion parameters 
         $ \delta $. The experimental crystal structure corresponds to 
         $ \delta = 0.0154 $.}
\label{tab1}
\end{table}
which comprises spin-polarized calculations with both ferro- 
and antiferromagnetic order. Obviously, the total energies for 
both situations assume a minimum for the experimental value of 
$ \delta $. However, for the antiferromagnetic solutions total 
energies are by $ \approx 27 $\,meV above those for the 
ferromagnetic case. In contrast, the optical band gap for both 
kinds of spin-polarized calculations assumes a maximum value at 
$ \delta = 0.0154 $; it vanishes for ferromagnetic order in the 
ideal $ {\rm K_2NiF_4} $ structure as has been also observed in 
our previous work on $ {\rm K_2CuF_4} $ \cite{eyert93}. 
Surprisingly, the magnetic moments at the Cr sites are almost 
independent of the distortion parameter $ \delta $, i.e.\ of 
the Cr-Cl bond lengths. This is due to the localized nature 
of these moments, which forms the basis for the treatment within 
the Heisenberg model, and reflects the finding of Feldkemper and 
Weber that a reasonable variation of hopping parameters changes 
the magnetic coupling only slightly \cite{feldkemper98}. 

\begin{figure}[htb]
\twoimages[width=65mm]{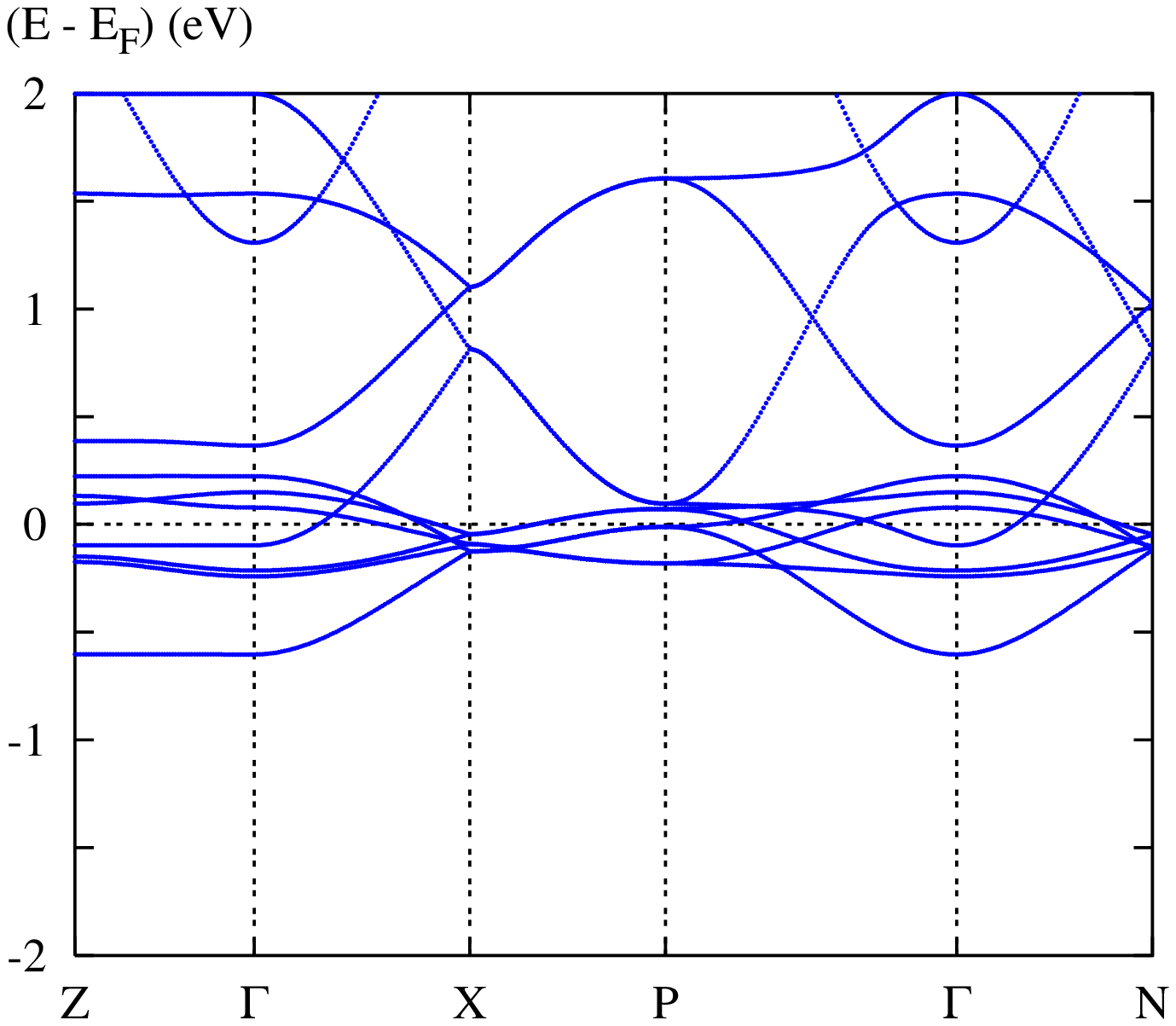}{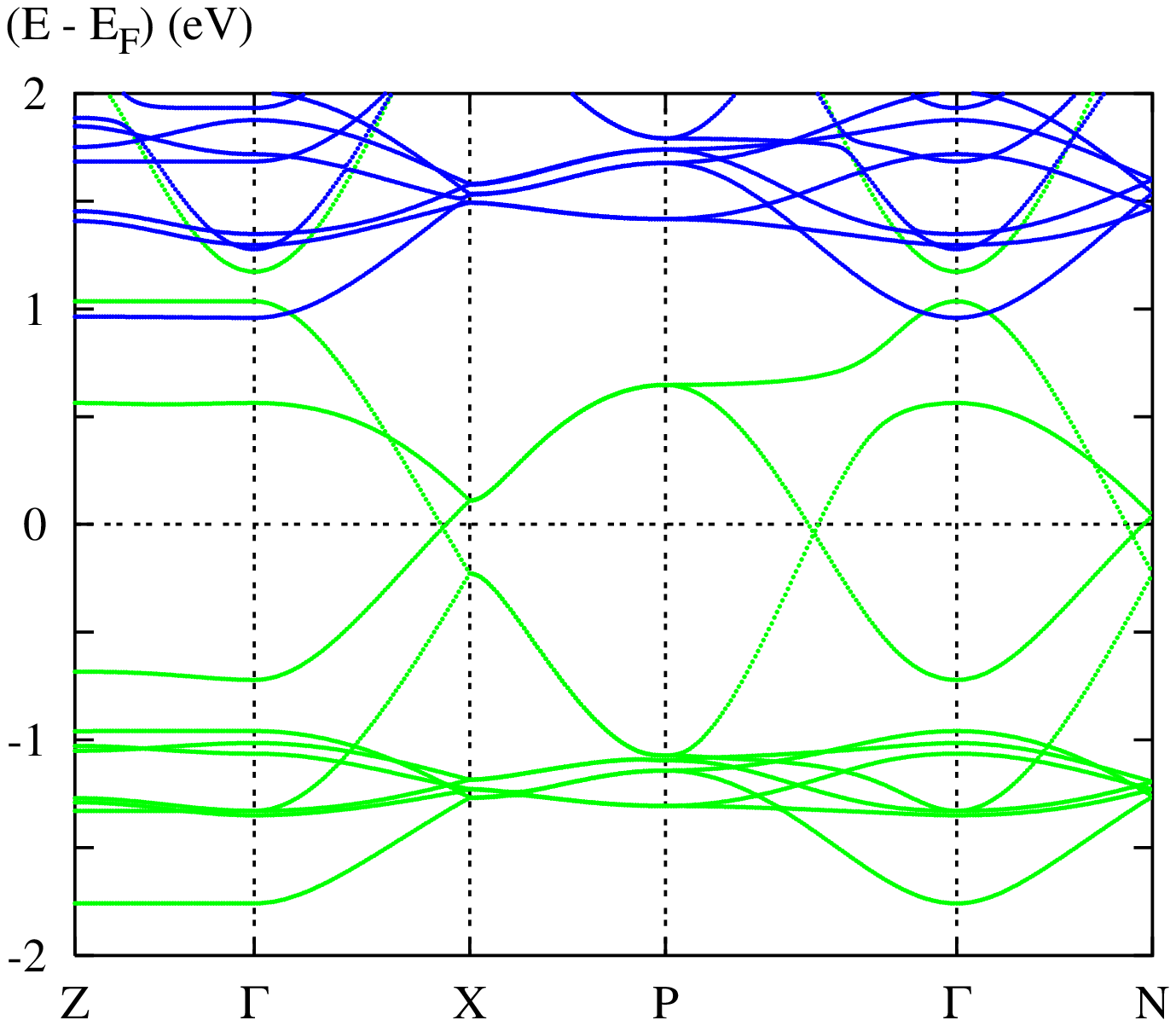}
\caption{Electronic band structure for the idealized tetragonal crystal 
         structure (distortion $ \delta=0 $).}
\label{fig4}
\end{figure}

Finally, we aim at understanding the role of the Jahn-Teller effect 
for the changes of the electronic states. To this end, we display 
band structures calculated for hypothetical crystal structures with 
zero distortion parameter $ \delta $ in Fig.\ 
\ref{fig4}.  
On comparing these results to Fig.\ \ref{fig2} we recognize as 
the main effect of the Jahn-Teller distortion the splitting of 
the bands in the middle of the line P-$ \Gamma $ as well as 
near the X- and N-points. While in the spin-degenerate case this 
splitting occurs at about 1\,eV above the Fermi energy, it opens 
the optical band gap for the ferromagnetically ordered state. 
Surprisingly, the spin-majority bands for $ \delta = 0 $ and 
finite distortion as displayed in these figures look very 
similar to the spin-degenerate bands of $ {\rm K_2NiF_4} $, 
where the band degeneracies occur exactly at the X- and N-points 
and thus give rise to perfect Fermi-surface nesting \cite{eyert93}. 
While the nesting is not complete here, one would nevertheless be 
tempted to interpret the Jahn-Teller effect as resulting from a 
Fermi-surface instability of the spin-majority bands. However, 
we point out that, due to the comparatively low Curie temperature 
of $ {\rm Rb_2CrCl_4} $, we cannot use the spin-polarized calculation 
as a starting point but would rather need the band structure of a 
paramagnetic system, which is not available.

\section{Conclusion}
In conclusion, we have reported on the first density functional based 
electronic structure calculation for the ferromagnetic semiconductor 
$ {\rm Rb_2CrCl_4} $. In agreement with the available experimental 
data we obtain a two-dimensional disperion of the electronic bands 
and a non-metallic ferromagnetically ordered ground state. The well 
localized magnetic moments of $ \approx 3.7 \mu_B $ are carried by 
the Cr $ 3d $ $ t_{2g} $ and $ d_{3z^2-r^2} $ orbitals, in perfect 
agreement with the magnetic moments and orbital occupations deduced 
from neutron diffraction experiments. Studying the influence of the 
Jahn-Teller distortion we find only minor influence on the size and 
ferromagnetic order of the magnetic moments, thus confirming the model 
approach by Feldkemper and Weber. In contrast, the electronic band 
structure experiences drastic changes on switching on the Jahn-Teller 
distortion turning, in particular, from metallic to semiconducting.

\begin{acknowledgments}
Fruitful discussions with W.\ Weber are gratefully acknowledged. 
This work was supported by the Deutsche Forschungsgemeinschaft 
through SFB 484. 
\end{acknowledgments}


\begin{thebibliography}{99}

\bibitem{tsubokawa60}
\Name{Tsubokawa I.}
\REVIEW{J.\ Phys.\ Soc.\ Jap.}{15}{1960}{1664}

\bibitem{rogado05}
\Name{Rogado N.\ S., Li, J., Sleight A.\ W., \and Subramanian, M.\ A.}
\REVIEW{Adv.\ Mater.}{17}{2005}{2225}

\bibitem{matar05}
\Name{Matar S.\ F., Subramanian, M.\ A., Eyert V., Whangbo M., \and 
      Villesuzanne A.}
\Review{cond-mat/0509431}

\bibitem{gregson75}
\Name{Gregson, A.\ K., Day P., Leech D.\ H., Fair M.J., \and Gardner W.\ E.} 
\REVIEW{J.\ Chem.\ Soc.\ (Dalton)}{}{1975}{1306}

\bibitem{hutchings81}
\Name{Hutchings M.\ T., Als-Nielsen J., Lindgard P.\ A., \and Walker P.\ J.}
\REVIEW{J.\ Phys.\ C:\ Solid State Phys.}{14}{1981}{5327}

\bibitem{hutchings76}
\Name{Hutchings M.\ T., Fair M.\ J., Day P., \and Walker P.\ J.}
\REVIEW{J.\ Phys.\ C:\ Solid State Phys.}{9}{1976}{L55}

\bibitem{janke83}
\Name{Janke E., Hutchings M.\ T., Day P., \and Walker P.\ J.}
\REVIEW{J.\ Phys.\ C:\ Solid State Phys.}{16}{1983}{5959}

\bibitem{kleemann86}
\Name{Kleemann W., Otte D., Usadel K.\ D., \and Brieskorn G.} 
\REVIEW{J.\ Phys.\ C:\ Solid State Phys.}{19}{1986}{395}

\bibitem{alsnielsen93}
\Name{Als-Nielsen J., Bramwell S.\ T., Hutchings M.\ T., McIntyre G.\ J., 
      \and Visser D.}
\REVIEW{J.\ Phys.:\ Condens.\ Matt.}{5}{1993}{7871}

\bibitem{bramwell95}
\Name{Bramwell S.\ T., Holdsworth P.\ C.\ W., \and Hutchings M.\ T.}
\REVIEW{J.\ Phys.\ Soc.\ Japan}{64}{1995}{3066}

\bibitem{khomskii73}
\Name{Khomskii D.\ I.\ \and Kugel K.\ I.}
\REVIEW{Solid State Commun.}{13}{1973}{763}

\bibitem{kugel82}
\Name{Kugel K.\ I.\ \and Khomskii D.\ I.}
\REVIEW{Sov.\ Phys.\ Usp.}{25}{1982}{231}

\bibitem{ito76}
\Name{Ito Y., \and Akimitsu J.}
\REVIEW{J.\ Phys.\ Soc.\ Japan}{40}{1976}{1333}

\bibitem{hidaka79}
\Name{Hidaka M., \and Walker P.\ J.}
\REVIEW{Solid State Commun.}{31}{1979}{383}

\bibitem{ledang77}
\Name{Le Dang K., Veillet P., \and Walker P.\ J.}
\REVIEW{J.\ Phys.\ C:\ Solid State Phys.}{10}{1977}{4593}

\bibitem{day79}
\Name{Day P., Hutchings M.\ T., Janke E., \and Walker P.\ J.}
\REVIEW{J.\ Chem.\ Soc.\ Chem.\ Commun.}{}{1979}{711}

\bibitem{muenninghoff80}
\Name{M\"unninghoff G., Treutmann W., Hellner E., Heger G., \and Reinen D.} 
\REVIEW{J.\ Solid State Chem.}{34}{1980}{289}

\bibitem{feldkemper98}
\Name{Feldkemper S.\ \and Weber W.}
\REVIEW{Phys.\ Rev.\ B}{57}{1998}{7755}

\bibitem{kuebler88}
\Name{K\"ubler J., Eyert V.\ \and Sticht J.}
\REVIEW{Physica C}{153-155}{1988}{1237}

\bibitem{eyert93}
\Name{Eyert V.\ \and H\"ock K.-H.}
\REVIEW{J.\ Phys.: Condens.\ Matt.}{5}{1993}{2987}

\bibitem{eyert95}
\Name{Eyert V., H\"ock K.-H., \and Riseborough P.\ S.}
\REVIEW{Europhys.\ Lett.}{31}{1995}{385}

\bibitem{eyert97}
\Name{Eyert V., Siberchicot B., \and Verdaguer M.}
\REVIEW{Phys.\ Rev.\ B}{56}{1997}{8959}

\bibitem{williams79}
\Name{Williams A.\ R., K\"ubler J., \and Gelatt C.\ D.\ jr.}
\REVIEW{Phys.\ Rev.\ B}{19}{1979}{6094}

\bibitem{eyert00b}
\Name{Eyert V.}
\REVIEW{Int.\ J.\ Quant.\ Chem.}{77}{2000}{1007}

\bibitem{eyert98}
\Name{Eyert V.\ \and H\"ock K.-H.}
\REVIEW{Phys.\ Rev.\ B}{57}{1998}{12727}

\bibitem{mixpap}
\Name{Eyert V.}
\REVIEW{J.\ Comp.\ Phys.}{124}{1996}{271}

\bibitem{fair77}
\Name{Fair M.\ J., Gregson A.\ K., Day P., \and Hutchings M.\ T.}
\REVIEW{Physica B}{86-88}{1977}{657}

\bibitem{muenninghoff82}
\Name{M\"unninghoff G., Hellner E., Fyne P.\ J., Day P., Hutchings M.\ T., 
      \and Tasset F.} 
\REVIEW{J.\ Phys.\ (France) Coll.\ C}{7}{1982}{243}

\end{thebibliography}
\end{document}